 \documentstyle[pre,aps,multicol,epsf,graphicx]{revtex}

\begin{document}

\draft

\title {Decay of DNLS breathers through inelastic multiphonon scattering}

\author{Magnus Johansson}

\address{Laboratoire L\'eon Brillouin (CEA-CNRS), 
   CEA Saclay, 
   F-91191 Gif-sur-Yvette Cedex, 
   France }

\date{\today}

\maketitle

%-----------------------------------------------------------------------
\begin{abstract} 
We consider the long-time evolution of weakly perturbed discrete nonlinear 
Schr\"odinger breathers. While breather growth can occur 
through nonlinear interaction with one single initial linear mode, breather 
decay is found to require excitation of at least two independent modes. 
All growth and decay processes of lowest order are found to disappear 
for breathers larger than a threshold value.
\end{abstract}


\vspace{-0.5cm}
\begin{multicols}{2}

%\narrowtext

%%%%%%%%%%%%%%%%%%%%%%%%%%%%%%%%%%%%%%%%%%%%%%%%%%%%%%%%%%%%%%%%%%%%%%%%

The study of intrinsically  localized modes in anharmonic lattices, {\em 
discrete breathers}, has yielded much attention during the last decade (see 
e.g.\ \cite{Aubry,Flach}). In particular, their existence as 
time-periodic solutions of nonlinear lattice-equations was proven 
under quite 
 general conditions \cite{MacKay}, and 
numerical schemes were developed for their explicit calculation 
 \cite{Marin96}. The generality of the concept of discrete breathers, 
which provide very efficient means to localize energy, 
has lead to numerous suggestions to its application in contexts 
where anharmonicity and discreteness are important, e.g.\ for 
describing energy and charge transport and storage in biological 
macromolecules \cite{PeyrardDNA}. Recently, discrete breathers have been 
experimentally observed in coupled optical waveguides \cite{Eisenberg}, 
in charge-density wave systems \cite{Swanson}, in magnetic systems 
\cite{Uli}, 
and in arrays of coupled Josephson junctions \cite{Josephson}.

Although a discrete breather under 
quite general conditions is linearly stable 
\cite{Aubry,MacKay,MarinCrete} (and thus no perturbations grow exponentially), 
there are many questions remaining concerning the long-time fate 
of perturbed breathers. 
In a previous paper \cite{growth}, some of these questions were addressed 
considering a particular model, the discrete nonlinear 
Schr\"odinger (DNLS) equation. The interaction between stationary breathers 
and  
small perturbations  corresponding to time-periodic eigensolutions to the 
linearized equations of motion around the breather was investigated 
using a multiscale perturbational approach, and it was found that the 
nonlinear interaction between the breather and single-mode small-amplitude 
perturbations could lead to breather {\em growth} through generation of 
radiating higher harmonics, but not to breather {\em decay}. It is the purpose 
of this Report to extend these results to more general perturbations of 
stationary DNLS breathers. In 
particular, we find that while the simplest growth process can be 
described as an inelastic scattering process with a one-frequency incoming 
wave and an 
additional outgoing higher-harmonic wave, the 
description of a decay process requires {\em at least two} 
incoming 
modes yielding outgoing modes with frequencies being linear combinations 
of the original ones.

In order to make this Report self-contained, we first recapitulate the 
main formalism from \cite{growth} (to which the reader 
is referred for details; see also \cite{PKA} for a similar approach in
continuous NLS models). With canonical 
conjugated variables $\{i \psi_n\},\{\psi_n^\ast\}$, the DNLS equation can be 
derived from the Hamiltonian
\begin{equation}
\label{DNLSham}
{\cal H} \left( \left\{i \psi_n\right\}, \left\{\psi_n^\ast \right\}\right) = 
\sum_n \left ( C |\psi_{n+1}- \psi_n |^2 - \frac{1}{2}| \psi_n|^4  \right ) 
\end{equation}
as
\begin{equation}
\label{DNLS}
i \dot{\psi_n} = \frac {\partial {\cal H }} {\partial \psi_n^\ast } 
= - C (\Delta \psi)_n - | \psi_n|^2 \psi_n ,
\end{equation}
where $\Delta$ is the discrete Laplacian, 
$(\Delta \psi)_n = \psi_{n+1} + \psi_{n-1} -2 \psi_n$, and we will assume 
$C>0$ without loss of generality. 
In addition to the Hamiltonian (\ref{DNLSham}), a second conserved quantity, 
the {\em excitation norm}, is defined as
\begin{equation}
\label{norm}
{\cal N } = \sum_n |\psi_n|^2.
\end{equation}
The conservation laws for these two quantities can be expressed in terms of 
continuity equations, with flux densities for the 
Hamiltonian and norm given by, respectively, 
%
%\begin{equation}
\begin{eqnarray}
\label{JHdef}
J_{\cal H} = - 2C {\rm Re} \left[\dot{\psi}_{n+1}(\psi_{n+1}^\ast 
- \psi_{n}^\ast )\right] ,
\\
%
%\end{equation}
%and
%\begin{equation}
\label{JNdef}
J_{\cal N}  = 2C  {\rm Im} 
\left[\psi_n^\ast \psi_{n+1} \right ].
%\end{equation}
\end{eqnarray}

The single-site breather is a localized stationary  solution to 
Eq.\ (\ref{DNLS}) of the form
$\psi_n (t)= \phi_n (\Lambda) e^{i \Lambda t}$,
where $\{\phi_n\}$ is time-independent with a single 
maximum {\em at} a lattice site (see e.g.\ \cite{Scott,MacKay,Weinstein}). 
It exists for all  $\Lambda/C > 0 $, and 
is a ground state solution minimizing 
the Hamiltonian (\ref{DNLSham}) for a fixed value of the norm (\ref{norm})
(see e.g. \cite{Weinstein}). Denoting the values of these quantities for the 
breather as ${\cal H}_\phi (\Lambda)$ ($< 0$) and ${\cal N}_\phi (\Lambda)$ 
 ($> 0$), 
respectively, we have: 
\begin{equation}
\frac {{\rm d}{\cal N}_\phi } { {\rm d } \Lambda} =  - \frac {1}{\Lambda}
\frac {{\rm d}{\cal H}_\phi } { {\rm d } \Lambda} > 0,
\label{NH}
\end{equation}
proving linear stability of the 
breather \cite{Laedke} as well as Lyapunov stability for 
norm-conserving perturbations \cite{Weinstein}.

To describe the dynamics close to the breather, we introduce the perturbation 
expansion (cf. \cite{growth})
\begin{eqnarray}
\psi_n(t)=\left\{\phi_n+\lambda \epsilon_n ( t ) + \lambda ^2 \eta_n (t) 
+ \lambda^3 \xi_n (t) \right. \nonumber\\
\left.+ \lambda^4 \theta_n (t) +...\right\} 
e ^ {i \int \Lambda dt } ,
\label{perturbation}
\end{eqnarray}
where  $\epsilon_n (0)$ is the initial small-amplitude perturbation. 
Substituting into Eq.~(\ref{DNLS}) and identifying coefficients for 
consecutive powers of the small parameter $\lambda$ yields for 
$\lambda^0, \lambda^1, \lambda^2, \lambda^3$, etc: 
\begin{eqnarray}
- \Lambda \phi_n + C (\Delta \phi)_n+|\phi_n|^2 \phi_n  
=  0,
\label{0th}\\
({\cal L }^{( \Lambda)} \epsilon)_n = 0, 
\label{1st}\\
({\cal L }^{(\Lambda)} \eta)_n =  
- \phi_n^\ast \epsilon_n^2 - 2 \phi_n | \epsilon _n |^2, 
\label{2nd}\\
({\cal L }^{(\Lambda)}\xi)_n =  
-2 \phi_n^\ast \epsilon_n \eta_n 
-4 \phi_n {\rm Re} \left[\epsilon_n \eta_n^\ast \right]
 - | \epsilon_n|^2 \epsilon_n ,
\label{3rd}
\end{eqnarray}
etc., where $({\cal L }^{( \Lambda) }\epsilon)_n \equiv 
i \dot{\epsilon}_n+ C (\Delta \epsilon)_n
+ 2 | \phi_n|^2 \epsilon_n  
 + \phi_n^2 \epsilon_n^\ast - \Lambda \epsilon_n$.  The 0th 
order equation (\ref{0th}) yields the breather 
shape $\{\phi_n\}$ (which for the single-site breather can be assumed real 
and positive without loss of generality), the 1st order equation 
(\ref{1st}) is the linearization of the DNLS equation for small breather 
perturbations, while the higher order equations describe the long-time 
dynamics on increasingly longer time-scales. 

The linearized solutions  are conveniently 
obtained by substituting 
$\epsilon_n(t) = \frac{1}{2} 
a\left (U_n+W_n \right) e ^{-i \omega t } + \frac{1}{2} a ^\ast 
\left (U_n^\ast-W_n^\ast \right) e^{i \omega t} $ 
into Eq.\ (\ref{1st}).  This yields (for real $\phi_n$) an eigenvalue problem 
of the form $\left ( \begin{array}{c} 0 \;\; \; {\cal L}_0 \\ {\cal L}_1 
\;\; \; 
0  \end{array} \right) \left ( \begin{array}{c} \{U_n\}\\\{W_n\} \end{array} 
\right)  = \omega \left ( \begin{array}{c} \{U_n\}\\\{W_n\} \end{array} 
\right)$, with Hermitian operators ${\cal L }_0$ and ${\cal L }_1$ defined by 
(cf.\ \cite{Carr})

\begin{eqnarray}
{\cal L }_0 W_n & \equiv  & -C ( \Delta W )_n  -\phi_n^2 W_n + 
\Lambda W_n ,
\label{W_n}\\
{\cal L }_1 U_n & \equiv  & -C ( \Delta U )_n - 3 \phi_n^2 U_n + 
\Lambda U_n. 
\label{U_n}
\end{eqnarray}
As the breather is linearly stable for all $\Lambda$ \cite{Laedke}, the
 eigenvalues $\omega$ are real, and the eigenvectors 
$\left( \{U_n\},\{W_n\}\right)$ can be chosen real and 
normalized.  The continuous (phonon) spectrum of extended 
eigensolutions is obtained for $|n|\rightarrow \infty$ 
($\phi_n \rightarrow 0$) yielding two uncoupled equations for 
the linear combinations $U_n \pm W_n$. 
Considering without loss of generality only the positive-frequency 
solutions $U_n + W_n\sim
 e ^{i q n }$ yields the dispersion relation
$\omega  = \Lambda +  2 C (1 - \cos q )$,  
so that  the continuous spectrum for $\omega>0$ is the interval 
$\omega\in [\Lambda,\Lambda+4C]$. 
Isolated eigenvalues $\omega \neq 0 $ outside the continuous spectrum give 
localized eigensolutions corresponding to breather internal 
modes \cite{JohanssonDresden,KPCP}: one spatially 
symmetric, 'breathing', mode exists for $0<\Lambda/C \lesssim 1.7$, 
and one antisymmetric, 'translational' or 'pinning' mode
for $0<\Lambda/C \lesssim 1.1$. (The variation of their 
frequencies with $\Lambda/C$ was shown in 
Fig.\ 1  of \cite{growth}). Finally, there are also 
the zero-frequency solutions to Eq.\ (\ref{1st}), obtained from 
the ansatz $\epsilon_n=U_n+i W_n$.  They can be written as a 
superposition of two fundamental modes: the 'phase mode'
\cite{AubryCretegny} $W_n=\phi_n$ describing 
a rotation of the overall phase of the breather, and the 'growth mode' 
\cite{AubryCretegny} $U_n=\partial \phi_n / \partial \Lambda $
(yielding $\epsilon_n=\partial \phi_n / \partial \Lambda + i \phi_n t $)
 describing a change of  breather frequency. 

This set of eigenvectors (including the zero-frequency modes) forms a basis 
for the space of solutions to Eq.~(\ref{1st}),  in which an 
arbitrary initial perturbation  can be expanded. 
Eigenvectors $\left( \{U_n^{(i)}\},\{W_n^{(i)}\}\right )$
with different (real) eigenvalues $\omega^{(i)}$ fulfill the 
orthogonality relations
\begin{equation}
\left(\omega^{(i)}-\omega^{(j)}\right)\sum_n \left ( 
U_n^{(i)}W_n^{(j)^\ast}+ W_n^{(i)}U_n^{(j)^\ast} \right) = 0 ,
\label{orthogonal}
\end{equation}
and, with the 'pseudoscalar' product defined from (\ref{orthogonal}), 
 the only nonzero product involving the zero-frequency modes is their 
cross-product \cite{Laedke}:
\begin{equation}
\sum_n \phi_n \frac {\partial \phi_n } { \partial \Lambda } = \frac {1} {2} 
\frac {{\rm d}{\cal N}_\phi } { {\rm d } \Lambda} > 0 .
\label{overlap}
\end{equation}

Let us now discuss the long-time effects of a small initial 
breather perturbation, expanded in the above basis. The second-order 
correction to the linearized dynamics is given 
by the inhomogeneous Eq.~(\ref{2nd}). As its right-hand side is quadratic in 
$\epsilon$, it describes  fundamental scattering processes 
involving 
at most two modes: assuming the initial perturbation $\epsilon_n(0)$ to 
consist of a set of modes with frequencies $\{\omega^{(i)}\}$, the right-hand side 
will contain the frequencies $\{2 \omega^{(i)}\}$, $\{\omega^{(i)}+\omega^{(j)}\}$,
$\{|\omega^{(i)}-\omega^{(j)}|\}$ and 0.
Thus,  it acts as a multi-periodic force localized at the breather 
region (since all terms are multiplied by $\phi_n$), and resonances with 
solutions to the homogeneous Eq.\ (\ref{1st}) typically yields a 
response for $\{\eta_n\}$ which either diverges in time (resonance with 
discrete spectrum) or is spatially unbounded (resonance with continuous 
part). 

In a standard way, the divergent parts can be removed by allowing 
for a slow, adiabatic time-evolution of the breather parameters. Similarly as 
in  \cite{growth}, where the case of 
single-mode initial perturbations was analyzed in detail, the divergence 
due to overlap 
between the static part of the right-hand side of (\ref{2nd}) and the 
zero-frequency modes can be shown to be equivalent to  a 
time-independent shift of the breather frequency $\Lambda$. Writing 
$\Lambda=\Lambda_0+\lambda^2 \Lambda_2$, where $\Lambda_0$ is the unperturbed 
breather frequency, the second-order shift $\Lambda_2$ will 
be the sum of the shifts resulting from the individual modes contained 
in $\epsilon_n(0)$. The latter was calculated in  \cite{growth} 
(cf.\ Eq.\ (31)),  and generalizes for the multimode case to
%\begin{equation}
%
$$
\Lambda_2=\frac{1}{\frac {{\rm d}{\cal N}_\phi } { {\rm d } \Lambda}} 
\sum_n \phi_n \frac {\partial \phi_n } { \partial \Lambda } 
\sum_i|a^{(i)}|^2
\left(3(U_n^{(i)})^2+(W_n^{(i)})^2\right), 
$$
%\label{static}
%\end{equation}
%
where $a^{(i)}$ is the initial amplitude for the mode 
$\left( \{U_n^{(i)}\},\{W_n^{(i)}\}\right )$ with frequency $\omega^{(i)}$. 
In general, $\Lambda_2$ can be either positive or negative, depending on the 
detailed spatial structure of the excited eigenmodes. 

The spatially unbounded response resulting from resonances between the 
oscillating part of the right-hand side of (\ref{2nd}) and phonon 
modes physically corresponds to emission of radiation from 
the breather region for each oscillation 
frequency belonging to the continuous spectrum of (\ref{1st}). 
The strength of the radiation fields can be calculated as 
illustrated in \cite{growth} for  second-harmonic 
($2 \omega^{(i)}$) resonances
(cf.\ Eq.\ (34) and the following discussion); here we give the corresponding 
results for two-mode ($\omega^{(i)} \pm \omega^{(j)} \equiv \omega^{(\pm)}$) 
resonances (assuming 
$\omega^{(i)}>\omega^{(j)}$ without loss of generality). 
Writing the corresponding responses as 
$\eta_n^{(\pm)}= \frac{1}{2}A_{ij}^{(\pm)} 
\left( u_n^{(\pm)}+w_n^{(\pm)} \right) e^{- i 
\omega^{(\pm)} t} +\frac{1}{2}A_{ij}^{(\pm)^\ast} 
\left( u_n^{(\pm)^\ast}-w_n^{(\pm)^\ast} 
\right) e^{i \omega^{(\pm)} t}$, with $A_{ij}^{(+)} = a^{(i)} a^{(j)}$ 
and $A_{ij}^{(-)} = a^{(i)} a^{(j)^\ast}$, 
the functions $u_n^{(\pm)}$ and $w_n^{(\pm)}$ are determined by
%
%\begin{equation}
%\label{M2omegap}
$\left ( \begin{array}{l} -\omega^{(\pm)} \;\; \;\;
{\cal L}_0 \\\; {\cal L}_1 \;\; -\omega^{(\pm)}  \end{array} \right) 
\left ( \begin{array}{c} \{u_n^{(\pm)}\} \\\{
w_n^{(\pm)}\} \end{array} \right)  = \frac{\phi_n}{2} \left ( \begin{array}{c} 
\{W_n^{(i)}U_n^{(j)}\pm U_n^{(i)} W_n^{(j)}\}\\\{3 U_n^{(i)}U_n^{(j)}
\mp W_n^{(i)}W_n^{(j)}\} \end{array} \right) .
$
%\end{equation}
When $\omega^{(\pm)}$ belongs to the phonon band, the 
radiation field strength will be  proportional to the overlap between 
 this right-hand-side and the corresponding continuous spectrum eigenvector 
$\left( \{U_n^{(\pm)}\},\{W_n^{(\pm)}\}\right )$, which using 
(\ref{orthogonal}) is obtained  as 
\begin{eqnarray}
\label{strength}
c^{(\pm)}=\frac{1}{2}\sum_n \phi_n \left[ W_n^{(\pm)}
\left( W_n^{(i)}U_n^{(j)}\pm U_n^{(i)} W_n^{(j)} \right) \right. \nonumber \\
\left. +U_n^{(\pm)} \left(3 U_n^{(i)}U_n^{(j)}
\mp W_n^{(i)}W_n^{(j)} \right) \right] . 
\end{eqnarray}
Far away from the breather, the radiation field should correspond to two 
identical outgoing propagating linear 
waves, yielding the boundary conditions
$ u_n^{(\pm)},w_n^{(\pm)} \rightarrow r^{(\pm)} e ^ {\pm i q^{(\pm)} n }, 
n \rightarrow \pm \infty $
where $r^{(\pm)}\sim c^{(\pm)}$ and  $q^{(\pm)} = 
\arccos \left ( 1 - \frac {\omega^{(\pm)} -\Lambda}{2C} \right)$ from 
the linear dispersion relation. 

Let us now discuss the consequences of the second-order radiation 
arising 
from the two-mode-interaction for the breather itself. 
 Assume for simplicity that only two modes with frequencies 
$\omega^{(1)}$ and $\omega^{(2)}$ ($\omega^{(1)}>\omega^{(2)}$) are 
initially excited (the presence of other modes will contribute to orders 
3 and higher), and that both modes belong to the continuous spectrum 
(the case of internal mode excitation requires some modification, see 
\cite{growth}, Sec. IIIC). Far away from the breather, we assume  a 
stationary regime which, in the most general case when $2\omega^{(1)}$, 
$2\omega^{(2)}$, $\omega^{(+)}$ and $\omega^{(-)}$ all belong to the 
phonon spectrum, corresponds to the boundary conditions at $n \rightarrow \pm 
\infty$: 
\begin{eqnarray}
\psi_n \rightarrow  e ^{i\Lambda t} \left[ a^{(1)} (e ^ {\mp i q^{(1)} n } + r^{(1)} 
e ^{\pm iq^{(1)} n })
e^ {-i \omega^{(1)} t }
\right.\nonumber \\
\left. + a^{(2)} (e ^ {\mp i q^{(2)} n } + r^{(2)} 
e ^{\pm iq^{(2)} n })
e^ {-i \omega^{(2)} t } \right.\nonumber\\
\left. + (a^{(1)})^2 r^{(1)}_2 e ^ { i (\pm q^{(1)}_2 n - 2 \omega^{(1)} t) } 
\right.\nonumber\\
\left.
+ (a^{(2)})^2 r^{(2)}_2 e ^ { i (\pm q^{(2)}_2 n - 2 \omega^{(2)} t) }
\right. \nonumber \\
\left. + a^{(1)}a^{(2)} r^{(+)} e ^ { i (\pm q^{(+)} n - \omega^{(+)} t) } 
\right.\nonumber\\
\left.+ a^{(1)}(a^{(2)})^\ast r^{(-)} e ^ { i (\pm q^{(-)} n - \omega^{(-)} t) } \right] .
\label{scattering}
\end{eqnarray}
Note that in general, the stationary amplitudes for the outgoing fundamental 
waves will differ from those of the incoming 
(i.e., $r^{(1)}, r^{(2)} \neq 1$); this is a consequence of 
resonances at the frequencies $\omega^{(1)}$ and $\omega^{(2)}$ in 
Eq.\ (\ref{3rd}) for the third-order field $\xi_n$. For a general 
multimode perturbation, it is seen from (\ref{3rd}) that this  correction 
takes the form $r^{(i)}=1+\sum_j \alpha_{ij} |a^{(j)}|^2$, where the sum goes 
over all initially excited modes (here $j=1,2$).

We then consider the conservation laws for the total norm and 
Hamiltonian, respectively, contained in some  region around the breather. 
Assuming the breather frequency $\Lambda$ to be the only time-dependent 
parameter in the stationary regime we can, similarly as in Ref.\ \cite{growth}, Sec.\ IV, 
write the time-averaged balance equations as 
\begin{equation}
\frac{{\rm d}\langle {\cal N } \rangle _t}{{\rm d}t} 
 = \frac {{\rm d}{\cal N}_\phi } { {\rm d } \Lambda} \dot{\Lambda} 
=\langle J_{\cal N}(-\infty)\rangle_t- \langle J_{\cal N}(+\infty)\rangle_t ,
\label{dNdtph}
\end{equation}
\begin{equation}
\frac{{\rm d}\langle {\cal H } \rangle _t}{{\rm d}t}  = 
\frac {{\rm d}{\cal H}_\phi } { {\rm d } \Lambda} \dot{\Lambda} 
=
\langle J_{\cal H}(-\infty)\rangle_t - \langle J_{\cal H}(+\infty)\rangle_t . 
\label{dHdtph}
\end{equation}
As the time-average of the flux densities $J_{\cal N}$ and $J_{\cal H}$ 
are additive quantities for small-amplitude plane waves 
of the form $\psi_n=A e^{i( Q n - \Omega  t)}$, the right-hand sides
are readily obtained from (\ref{scattering}) using the general expressions 
$J_{\cal N} = 2 |A|^2 C \sin Q$ and $J_{\cal H} = \Omega J_{\cal N}$ for 
the individual waves resulting from (\ref{JHdef})-(\ref{JNdef})
Combining Eqs.\ (\ref{dNdtph})-(\ref{dHdtph}) and using (\ref{NH}) yields  
the time-derivative of 
the breather frequency to order 4 in the mode amplitudes $a^{(1)}, a^{(2)}$ as:
\begin{eqnarray}
\dot{\Lambda}  = \frac { 4C}{\frac{{\rm d}{\cal N}_\phi }{ {\rm d } \Lambda} }
\left[|a^{(1)}|^4 |r_2^{(1)}|^2 \sin q_2^{(1)}\right.\nonumber\\
+ |a^{(2)}|^4\left(\frac{2\omega^{(2)}}{\omega^{(1)}}-1\right) |r_2^{(2)}|^2
\sin q_2^{(2)} \nonumber \\
+|a^{(1)}|^2|a^{(2)}|^2\frac{\omega^{(2)}}{\omega^{(1)}}
\left(|r^{(+)}|^2 \sin q^{(+)} - |r^{(-)}|^2 \sin q^{(-)} \right) \nonumber \\
\left. +|a^{(2)}|^2\left(1-\frac{\omega^{(2)}}{\omega^{(1)}}\right)
\left(1-|r^{(2)}|^2\right) 
\sin q^{(2)} \right] .
\label{Lambdadotph}
\end{eqnarray}
For a single-mode initial excitation, $a^{(2)}=0$ and only the first, positive 
term of the right-hand side remains, so that we recover the result (Eq.\ (55)  
in  \cite{growth}) that the generation of second-harmonic radiation always 
leads to breather {\em growth}. However, in the two-mode case, the expression 
(\ref{Lambdadotph}) is in general not sign-definite, and in particular 
the contribution from the frequency $\omega^{(-)}$ 
is always negative. It is therefore natural to associate this radiation 
with the {\em fundamental lowest-order mechanism for breather decay}. 
This can be seen as a consequence of the Hamiltonian flux 
density being proportional to the frequency for plane waves. 
Thus the scattering towards the lower frequency $\omega^{(-)}$
should yield a net flow of negative Hamiltonian energy out from the breather 
region, to which the breather adapts by decreasing its frequency and 
amplitude according to (\ref{NH}). On the other hand, the 
generation of radiation with the higher frequency 
$\omega^{(+)}$ should analogously contribute to 
breather growth, as for the second-harmonic case. 
(See also \cite{CAF} for a similar explanation
of breather decay in a Klein-Gordon model resulting from a resonance in the 
linearized equations.)

It is important to note, that when $\omega^{(1)}$ and  $\omega^{(2)}$ are 
phonon modes, they fulfill $\Lambda \leq \omega^{(1)}, \omega^{(2)} 
\leq \Lambda + 4C$, so that we have $0 \leq \omega^{(1)}- \omega^{(2)} \leq 
4C$ and $2 \Lambda \leq 2\omega^{(1)}, 2\omega^{(2)}, \omega^{(1)}+ \omega^{(2)}
\leq 2 \Lambda + 8 C$. Thus, we see that second order radiation can  only be 
generated if the breather frequency fulfils $0 < \Lambda \leq 4C$, since 
for $\Lambda > 4C$ the frequency  $\omega^{(-)}$ is always below the phonon 
band, while $2\omega^{(1)}, 2\omega^{(2)}$ and  $\omega^{(+)}$ are
always above. Therefore, when $\Lambda>4C$ all terms in the right-hand side of 
Eq.\ (\ref{Lambdadotph}) necessarily vanish, and {\em all growth and decay 
processes of lowest order disappear}. Numerically, we found \cite{growth} that 
this corresponds to breathers with central-site intensity $|\psi_{n_0}|^2 
\gtrsim 5.65$. Therefore, these large-amplitude breathers are particularly 
stable, since all possible growth and decay processes result from third- and 
higher-order radiation processes so that the rate of their growth/decay 
must be at least of order 6 in the initial mode amplitudes. 

\begin{figure}[ht]
\noindent
\includegraphics[height=8.8cm,angle=270]{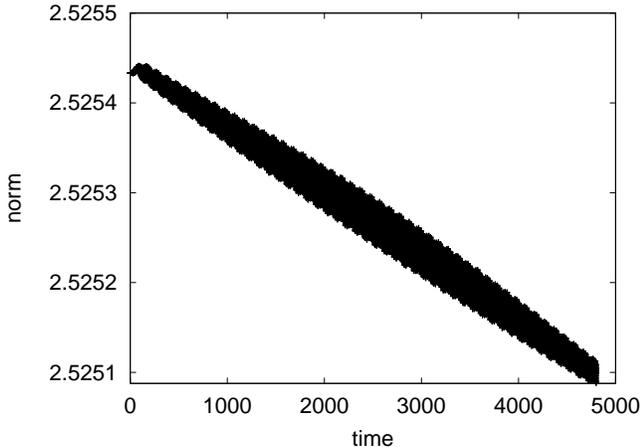}
\caption{Time evolution of the total norm ${\cal N}$ contained in a region of 
120 sites around a breather with frequency $\Lambda=0.45$, perturbed along 
two eigenmodes with frequencies $ \omega^{(1)} \approx 3.00$ and  
$\omega^{(2)} \approx 2.47$, respectively ($C=1$).
}
\label{fig1}
\end{figure}

To illustrate the fundamental lowest-order decay mechanism, we consider the 
'pure' case 
when $\omega^{(-)}$ belongs to the phonon spectrum, while $2 \omega^{(1)}, 
2\omega^{(2)}$, and $\omega^{(+)}$ are outside. Thus, we have 
$r_2^{(1)}=r_2^{(2)}=r^{(+)}=0$, and only the last two terms in 
Eq.\ (\ref{Lambdadotph}) are nonzero. We note that if $|r^{(2)}|<1$, the last 
term will be positive, and we can therefore in general not conclude that 
$\dot{\Lambda}$ must be negative. The reason is, that although the scattering 
towards the lower frequency $\omega^{(-)}$ from either of the two frequencies 
$ \omega^{(1)}$ or $\omega^{(2)}$ considered independently would yield a net 
outflow of negative Hamiltonian energy from the breather region and thereby 
breather decay, there will also be a contribution from the mixing between 
the $ \omega^{(1)}$ and  $\omega^{(2)}$ modes arising from the third order 
equation (\ref{3rd}). This contribution would yield a net outflow of positive 
Hamiltonian energy if $|r^{(2)}|<1$ and $|r^{(1)}|>1$, and could therefore 
contribute to breather growth. However, extensive numerical simulations 
for different $\Lambda$, $\omega^{(1)}$ and $\omega^{(2)}$ belonging to this 
region 
of 'pure' $\omega^{(-)}$-scattering have always shown that the net result is 
a linear {\em decrease} of the breather frequency,
and therefore we believe it justified to identify this 
two-wave scattering as the fundamental lowest-order breather decay mechanism 
(Fig.\ \ref{fig1}). 

\vspace{-0.5cm}
In conclusion, we found that while in the DNLS model breather growth 
can result through interaction between the breather and single-mode 
initial perturbations, 
the description of breather decay requires simultaneous excitation of at 
least 
two independent linear modes. This confirms numerical results  
in Ref.\ \cite{growth} (Figs.\ 5 and 6), showing breather 
decay also from initially single-mode perturbations with 
larger amplitude, as more frequencies became gradually excited 
e.g.\ through oscillatory wave instabilities \cite{Anna}. We believe that our 
approach could be useful to understand the properties of the 
stationary intensity probability distribution function in the 'negative 
temperature'-regime of the DNLS model \cite{Kim},  
where persistent localized 
breathers were found, weakly interacting with small amplitude 
radiation. Finally, we remark that the DNLS model is nongeneric among 
Hamiltonian lattice models as it has two conserved quantities, 
and the DNLS  
breather has only one fundamental frequency with no harmonics. Thus, 
our approach cannot immediately be extended to other models 
exhibiting breathers such as Fermi-Pasta-Ulam or Klein-Gordon lattices. 
However, as for the latter the DNLS equation is known 
(see e.g.\  \cite{Anna}) to describe the small-amplitude dynamics 
for small inter-site coupling, we believe that the 
breather growth and decay mechanisms described here are relevant also in 
these systems. This  will be investigated in future work. 

I am grateful to S.\ Aubry for initiating my interest in this problem
and for valuable discussions, as well as to G.\ Kopidakis and 
K.\ \O.\ Rasmussen for useful remarks.  An EC Marie Curie
fellowship is gratefully acknowledged.

%%%%%%%%%%%%%%%%%%%%%%%%%%%%%%%%%%%

\end{multicols}
\end{document}